\documentclass[conference,a4paper]{IEEEtran} 
\pdfoutput=1

\IEEEoverridecommandlockouts

\usepackage{graphicx,cite}
\usepackage{amsmath}
\usepackage{url}

\usepackage{amssymb}
\usepackage{amsfonts}
\usepackage{lineno}
\usepackage{dsfont}
\usepackage{bbm}
\usepackage{times}
\usepackage{algorithm}
\usepackage{algcompatible}

\newcommand*{\myfigfactor}{0.4}

\DeclareMathOperator*{\argmax}{arg\,max}

\newcommand{\vect}{\boldsymbol}

\newcommand{\vectab}[2]{#1_1,\ldots,#1_{#2}}

\newcommand{\Seta}[1]{1,\ldots,#1}
\newcommand{\Set}[1]{\mathcal{#1}}

\newcommand{\vectx}{\vect{x}}
\newcommand{\vecty}{\vect{y}}

\newcommand{\texth}[1]{\!^\text{#1}}

\usepackage{forloop}
\newcounter{loopcntr}
\newcommand{\rpt}[2][1]{%
	\forloop{loopcntr}{0}{\value{loopcntr}<#1}{#2}%
}

\newcommand{\subgroup}[1]%
{\rlap{\smash{%
	\newcount\cnt%
	\cnt \numexpr#1\relax%
	\advance\cnt -1\relax%
	$\tabcolsep=.1em\begin{tabular}[t]{|l}\multicolumn{1}{l}{}\\%
	\rpt[\cnt]{\\}
	\\\hline\end{tabular}$%
}}}

\begin{document}

\title{\huge Dynamic Clustering and Sleep Mode Strategies \\ for Small Cell Networks
\vspace{-18pt}
}

\author{
\IEEEauthorblockN{Sumudu Samarakoon\IEEEauthorrefmark{1}, Mehdi Bennis\IEEEauthorrefmark{1}, Walid Saad\IEEEauthorrefmark{2} and Matti Latva-aho\IEEEauthorrefmark{1} \\}
\IEEEauthorblockA{\small\IEEEauthorrefmark{1}Centre for Wireless Communications, University of Oulu, Finland, email: \{sumudu,bennis,matti.latva-aho\}@ee.oulu.fi \\
\IEEEauthorrefmark{2}Wireless@VT, Bradley Department of Electrical and Computer Engineering, Virginia Tech, Blacksburg, VA, email: walids@vt.edu}
\vspace{-31pt}
\thanks{This work is supported by the SHARING project under the Finland grant 128010 and the U.S. National Science Foundation (NSF) under Grants CNS-1253731 and CNS-1406947.}
}

\IEEEpubid{\makebox[\columnwidth]{978-1-4799-5863-4/14/\$31.00~\copyright~2014 IEEE \hfill}\hspace{\columnsep}\makebox[\columnwidth]{}}

\maketitle
\nopagebreak[4]
\begin{abstract}

In this paper, a novel cluster-based approach for optimizing the energy efficiency of wireless small cell networks is proposed.
A dynamic mechanism based on the spectral clustering technique is proposed to dynamically form clusters of small cell base stations.
Such clustering enables intra-cluster coordination among the base stations for optimizing the downlink performance through load balancing, while satisfying users' quality-of-service requirements. 
In the proposed approach, the clusters use an opportunistic base station sleep-wake switching mechanism to strike a balance between delay and energy consumption.
The inter-cluster interference affects the performance of the clusters and their choices of active or sleep state.
Due to the lack of inter-cluster communications, the clusters have to compete with each other to make decisions on improving the energy efficiency.
This competition is formulated as a noncooperative game among the clusters that seek to minimize a cost function which captures the tradeoff between energy expenditure and load.
To solve this game, a distributed learning algorithm is proposed using which the clusters autonomously choose their optimal transmission strategies. 
Simulation results show that the proposed approach yields significant performance gains in terms of reduced energy expenditures up to $40\%$ and reduced load up to $23\%$ compared to conventional approaches.

\end{abstract}

\begin{keywords}
	Clustering, energy efficiency, game theory, small cell networks
\end{keywords}
\vspace{-12pt}
\section{Introduction}\label{sec:introduction}

The emergence of bandwidth-intensive wireless services has strained current cellular systems and led to an increased energy consumption in wireless communications systems~\cite{pap:brevis11}.
In this respect, the development of energy-efficient resource management mechanisms for wireless networks has become a major research interest in recent years. 
In particular, the deployment of low-cost, low-power small cells over existing cellular networks has been introduced as a promising solution for providing energy-efficient wireless resource management~\cite{jnl:amr14}.

Existing literature has studied a number of problems related to resource allocation in small cell networks (SCNs) including load balancing, power control, and base station (BS) sleep-wake mechanisms, among others \cite{pap:brevis11,pap:bhuaumik10,pap:soh13,jnl:amr14,pap:hoisseini12,pap:lee13}.
However, most of these existing solutions often rely on a central controller which gathers all network information and makes all decisions.
Such a centralized approach to SCN resource management introduces additional costs and overhead to the system.
Therefore, providing distributed, self-organizing capabilities to BSs has become a central research topic in recent years.
In this respect, different types of self-organizing sleep mode strategies that can be used by SCNs to optimize the energy consumption and the BS loads are proposed in \cite{pap:bhuaumik10} and \cite{pap:soh13}.
However, the authors in~\cite{pap:bhuaumik10} illustrate some practical challenges with such techniques that relate to the lack of inter-BS coordination, such as determining the efficient operating point, efficiently utilizing radio resources among active BSs, and avoiding outages due to the selfish behavior of BSs.
Developing a clustering approach using which the BSs can coordinate with minimal information exchange can be a suitable solution to the above challenges~\cite{jnl:luxburg07}. 
Multiple interference management techniques which improve the energy efficiency based on clustering are proposed in~\cite{jnl:amr14,pap:hoisseini12}, and~\cite{pap:lee13}.
Therein, clustering has been performed based on static BS locations.

The main contribution of this paper is to develop a dynamic cluster-based mechanism for switching BSs ON/OFF  
depending on parameters such as the current traffic load, energy expenditures, and the BS density.
Unlike previous studies~\cite{jnl:amr14,pap:hoisseini12}, and \cite{pap:lee13}, we propose a clustering approach that incorporates, beyond classical location-based metrics, the effects of the time-varying BS load.
Clustering enables intra-cluster coordination among the base stations for the purpose of optimizing the downlink performance via load balancing.
Within each cluster, the BSs adopt an opportunistic sleep-wake mechanism to reduce the energy consumption.
Due to inter-cluster interference, the clusters have to compete with one another to make decisions on improving the energy efficiency via a dynamic choice of sleep and active states.
We cast the problem of dynamic sleep state selection as a non-cooperative game between clusters of BSs.
To solve this game, we propose a distributed algorithm using notions from Gibbs-sampling~\cite{jnl:qian12}.
The proposed algorithm allows the SCN's BS to switch ON/OFF by offloading the traffic within the cluster, while satisfying users' quality-of-service requirements.  
Simulation results show that the proposed approach improves the SCN's energy efficiency and reduces the overall load in the system as compared to conventional approaches.

The rest of the paper is organized as follows. 
The system model and problem formulation are presented in Section~\ref{sec:system_model_problem_form}.  
The cluster-based coordination mechanism is presented in Section~\ref{sec:clusters}.
In Section~\ref{sec:solution}, the proposed game-theoretic approach and its corresponding learning algorithm are discussed.
Simulation results are presented and analyzed in Section~\ref{sec:results}. 
Finally, conclusions are drawn in Section~\ref{sec:conclusions}.

\section{System Model and Problem Formulation}\label{sec:system_model_problem_form}

\subsection{Network Model}\label{sec:network_model}

Consider the downlink transmission of a wireless network consisting of a set of small cell base stations (SBSs) $\Set{B}=\{\Seta{B}\}$
underlaid on a macro cellular network. 
We assume that the BSs are uniformly distributed over a two-dimensional network layout and we let $\vectx$ be any location on the two-dimensional plane measured with respect to the origin.
Moreover, let $\Set{L}_b$ be the coverage area of BS $b$ such that any given user equipment (UE) at a given location $\vectx$ is served by BS $b$ if $\vectx\in\Set{L}_b$.
Furthermore, we consider that all BSs are grouped into a set of clusters $\overline{\Set{C}}=\{\vectab{\Set C}{|\overline{\Set{C}}|}\}$ in which any cluster ${\Set C}\in\overline{\Set{C}}$ consists of a set of BSs who can cooperate with one another.
Note that $|\overline{\Set{C}}|$ denotes the cardinality of the set $\overline{\Set{C}}$.
Here, we assume that, within a given cluster ${\Set C}\in\overline{\Set{C}}$, the BSs allow to efficiently offload the UEs among each others and enable sleep mode while maintaining the UEs' quality of service.

Let $I_b$ be the transmission indicator of BS $b$ such that $I_b=1$ indicates the active state while $I_b=0$ reflects the idle or sleep state.
Here, we assume that, in active state, each BS will service all UEs in its coverage area.
From an energy efficiency perspective, some BSs might have an incentive to switch into sleep mode.
Note that during the idle state, a BS consumes nonzero power to sense the UEs in its vicinity. 
The power consumption of a BS $b$ is given by~\cite{jnl:kyuho11}:
\begin{equation}\label{eqn:state_based_power}
	P^{\texth{Total}}_b =
	\begin{cases}
		P_b^{\texth{Idle}} &\mbox{if}~I_b=0, \\
 		\big( P_b^{\texth{Work}} + q_bP_b^{\texth{Idle}} \big) &\mbox{if}~I_b=1,
	\end{cases}
\end{equation}
\noindent where $P_b^{\texth{Idle}}$ is the power consumption at the baseband unit during the sleep state and $P_b^{\texth{Work}}$ is the effective transmission power of BS $b$. 
$q_b > 1$ is a parameter that allows to capture the fact that, when a BS is switched to an active state, it has to turn on its power amplifier, radio frequency, and backhaul components which will constitute an increase of energy compared to the idle state~\cite{pap:frenger11}.

We assume that all BSs transmit on the same frequency spectrum (i.e., co-channel deployment) and a BS schedules one UE at a time.
Therefore, 
the data rate at a given location $\vectx\in\Set{L}_b$ from BS $b$ with transmission power $P_b$ is given by:
\begin{equation}\label{eqn:rate_ue}
	R_b(\vectx) = \omega\log_2\Big(1 + \frac{ P_bI_b h_{b}(\vectx) }{ \sum_{\forall b'\in\Set{B}/b} P^{\texth{Work}}_{b'}I_{b'}h_{b'}(\vectx) + N_0 } \Big),
\end{equation}
\noindent where $h_{b'}(\vectx)$ is the channel gain from BS $b'$ to a given UE location at $\vectx$, $N_0$ is the noise variance and $\omega$ is the bandwidth.

We assume that the UEs connected to BS $b$ are heterogeneous in nature such that each UE has a different QoS requirement based on its individual packet arrival rate.
In this respect, let $\lambda(\vectx)$ and $1/\mu(\vectx)$ be the packet arrival rate and the mean packet size of any UE at location $\vectx\in\Set{L}_b$.
The data rate offered to the UE at location $\vectx$ from BS $b$ is $R_b(\vectx)$ and thus, the load density of BS $b$ becomes $\vect{\varrho}_b = \{\varrho_b(\vectx)|\vectx\in\mathcal{L}_b\}$ where $\varrho_b(\vectx)=\frac{\lambda_b(\vectx)}{\mu_b(\vectx)R_b(\vectx)}$.
Here, the load density $\vect{\varrho}_b$ represents the fraction of time needed to transfer the traffic $\frac{\lambda_b(\vectx)}{\mu_b(\vectx)}$ from BS $b$ to location $\vectx$~\cite{jnl:hongseok12}.
Consequently, the BS load $\rho_b$ of BS $b$ is given by:
\begin{equation}\label{eqn:load_bs}
	\rho_b = \int_{\vectx\in\Set{L}_b} \varrho_b(\vectx) d\vectx. 
\end{equation}
\noindent
Due to the fact that the BS load $\rho_b$ reflects the fractional operation time, for a given BS transmision power $P_b$, the effective transmission power is  $P^{\texth{Work}}_b = \rho_b P_b$. 
\subsection{Problem Formulation}\label{sec:formulations}

The configuration of the entire network is defined by the transmission powers and the states of all the BSs.
This configuration can thus be captured by a transmission power vector $\vect{P}=(\vectab{P}{|\Set{B}|})$
and a state indicator vector $\vect{I}=(\vectab{I}{|\Set{B}|})$.
The network configuration $\vect{v} = (\vectab{v}{|\Set{B}|})$, where the configuration of BS $b$ is $v_b=(P_b,I_b)$, is coupled with the load vector $\vect{\rho}=(\vectab{\rho}{B})$ following (\ref{eqn:state_based_power})-(\ref{eqn:load_bs}).

For a given network configuration $\vect{\rho}$, BS $b$ handles the load $\rho_b$ from the set of UEs in its coverage area $\Set{L}_b$.
Each BS $b$ can increase or decrease the handled load $\rho_b$ by fine-tuning the energy consumption.
Thus, there is a tradeoff between load (delay) and energy consumption reduction.
Here, for each BS $b\in\Set{B}$, we define a cost function that captures both energy consumption and load, as follows:
\begin{equation}\label{eqn:bs_utility_metric}
	\Gamma_b(\vect{v}) = \Gamma_b(\vect{\rho}) = \alpha P_b^{\texth{Total}} \; + \; \beta \rho_b,
\end{equation}
where the coefficients $\alpha$ and $\beta$ are weight parameters that indicate the impact of energy and load on the cost, respectively.
For any cluster $\Set{C}\in\overline{\Set{C}}$, the cluster-wise cost $\Gamma_{\Set{C}}$ is the aggregated cost of each cluster member, i.e. $\Gamma_{\Set{C}}(\vect{\rho}) = \sum_{\forall b\in\Set{C}} \Gamma_b(\vect{\rho})$.
Our overall objective is to minimize the total network cost as given by the following optimization problem:
\begin{subequations}
\begin{eqnarray}\label{eqn:optimization_energy_efficiency_network}
	 \underset{\vect{\rho},\overline{\Set{C}}}{\text{minimize}} && \textstyle\sum_{\forall\Set{C}\in\overline{\Set{C}}} \Gamma_{\Set{C}}(\vect{\rho}) \\ 
		\label{cns:BI_belongs_to_cluster} \text{subject to} && |\Set{C}(b)|\geq 1, \quad \forall b\in\Set{B} \\
		 \label{cns:exclusive_clusters} && \Set{C}\cap\Set{C}'=\emptyset, \quad \forall \Set{C},\Set{C}'\in\overline{\Set{C}},~\Set{C}\neq\Set{C}' \\
		\label{cns:load} &&
			0 \leq \textstyle\sum_{\forall b\in\mathcal{C}}\rho_b \leq 1, \quad \forall \Set{C}\in\overline{\Set{C}} \\
			\label{cns:power} && P_b^{\texth{Total}} \leq P_b^{\texth{Max}}, \quad \forall b\in\Set{B} \\
			\label{cns:state} && I_b \in \{0,1\}, \quad \forall b\in\Set{B}
\end{eqnarray}
\end{subequations}
\noindent
where $P_b^{\texth{Max}}$ is the maximum transmit power of BS $b$.
Here, the constraints (\ref{cns:BI_belongs_to_cluster}) and (\ref{cns:exclusive_clusters}) ensure that any BS is part of only one cluster.

\section{Cluster Formation and In-Cluster Coordination}\label{sec:clusters}

\begin{figure*}
\centering
\includegraphics[width=.85\linewidth]{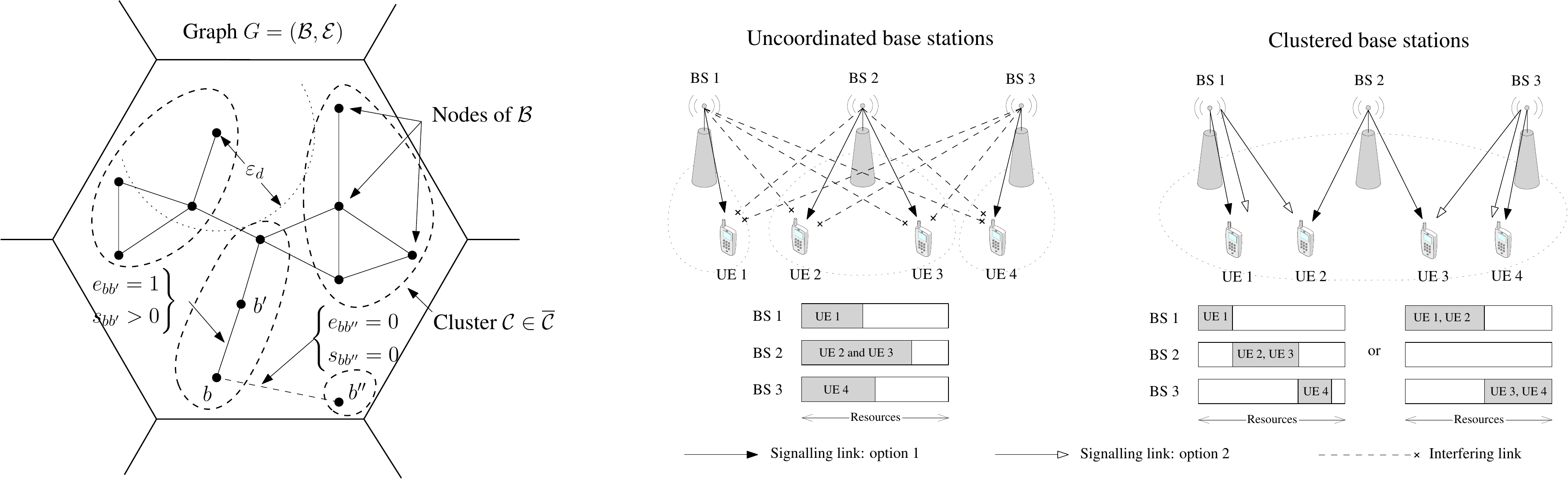}
\vspace{-10pt}
\caption{Graph representation of the network and the intra-cell interference mitigation through clustering based coordination. Clustered base stations do not interfere with each others as well as users may offloaded within the cluster members.}
\label{fig:coordination}
\vspace{-20pt}
\end{figure*}

For clustering, we map the system into a weighted graph $G=(\Set{B},\Set{E})$ as illustrated in Fig.~\ref{fig:coordination}.
Here, the set of BSs $\Set{B}$ represents the nodes while $\Set{E}$ represents the links between BSs.

\subsection{Cluster formation}

The key step in clustering is to identify similarities between network elements (i.e. BSs) in which
BSs with similar features can be  grouped.
This will allow to perform coordination between BSs with little signaling overhead.
Here, we propose a number of techniques to calculate the similarities between BSs based on their locations and loads.

\subsubsection{Adjacency-based neighborhood and Gaussian similarity}

Consider the graph $G=(\Set{B},\Set{E})$. 
The set of edges $\Set{E}$ indicates the adjacency between nodes.
Here, we use a parameter $e_{bb'}$ to characterize the presence of a link between nodes $b$ and $b'$ when $e_{bb'}=1$.
Let $\vecty_b$ represent coordinates of the vertex (or BS) $b\in\Set B$ in the Euclidean space.
The neighborhood of the node $b$, $\Set{N}_{b}=\{b'|~e_{bb'}=e_{b'b}=1, \|\vecty_b-\vecty_{b'}\|\leq\varepsilon_d \}$ is defined by the adjacency of the nodes in the $\varepsilon_d$-neighborhood. 

The links between the nodes are weighted based on their similarities.
Since nearby BSs are  more likely to cooperate,
we use the Gaussian similarity metric as a means to compute the weight between two nodes $b,b'\in\Set{B}$ as follows~\cite{jnl:luxburg07}:
\begin{equation}\label{eqn:gaussian_similarity}
	s_{bb'}^d =
	\begin{cases}
		\exp\big(\frac{-\|\vectx_b-\vectx_{b'}\|^2}{2\sigma_d^2}\big) & \mbox{if}~e_{bb'}=1,\\
		0 & \mbox{otherwise},
	\end{cases}
\end{equation}
where the parameter $\sigma_d$ controls the impact of the neighborhood size. 
Furthermore, the distance-based similarity matrix $\vect{S}^d$ is formed using $s^d_{bb'}$ as the $(b,b')$-th entry.

\subsubsection{Load-based dissimilarities}

Unlike the static distance-based clustering, load-based clustering provides a more dynamic  manner of grouping neighboring BSs in terms of traffic load.
Therefore, the load based similarity between BSs $b$ and $b'$ can be computed as follows;
\begin{equation}\label{eqn:similarity_load}
	s^{l}_{bb'} = \exp\bigg(\frac{\|\rho_b-\rho_{b'}\|^2}{2\sigma_l^2}\bigg),
\end{equation}
where $\sigma_l$ controls the range of the similarity. 
The value of $s^l_{bb'}$ is used as the $(b,b')$-th entry of the load based similarity matrix $\vect{S}^l$.

The typical method for combining similarities is by forming a linear combination such as $\big(\theta s_{bb'}^d + (1-\theta)s_{bb'}^l\big)$ with $0\leq\theta\leq 1$.
The main drawback of a linear combination is that
there is a possibility of having two BSs with positive similarity due to their loads although a physical link is impossible due to their locations.
Therefore, the joint similarity matrix $\vect{S}$ with $s_{bb'}$ as the $(b,b')$-th element is formulated as follows
\begin{equation}\label{eqn:similarity_joint}
	s_{bb'} = [s_{bb'}^d]^\theta \cdot [s_{bb'}^l]^{(1-\theta)},
\end{equation}
with $0\leq\theta\leq 1$ which controls the  impact of the distance and the load similarities on the joint similarity.
Here, the cooperation is possible only if a physical link between nodes exists, i.e. $\forall \theta\in[0,1]$, $e_{bb'}=0\implies s_{bb'}=0$.

For BS clustering we use a \emph{spectral clustering technique} which exploits the compactness and the connectivity of the nodes in a graph~\cite{jnl:luxburg07}.
Since the joint similarity in (\ref{eqn:similarity_joint}) defines how well the BSs are connected to each other, spectral clustering is seen as an appropriate technique.

First, the graph Laplacian matrix $\vect{L}$ is formed with the $(b,b')$-th element as $l_{bb'} = \sum_{b_k=1}^{|{\Set{B}}|} (s_{bb_k} - s_{bb'})$.
For a given cluster size $|\overline{\Set{C}}|$, the concatenation of the smallest $|\overline{\Set{C}}|$ eigenvectors of $\vect{L}$ can be used to build the modified similarity matrix $\tilde{\vect S}$.
Then, the $k$-mean clustering algorithm is applied on $\tilde{\vect S}$ in order to produce the clusters $\overline{\Set{C}}$.
The number of clusters $k$ is 
closely related to the eigenvalues of $\vect L$ and can be found as follows~\cite{jnl:luxburg07};
\begin{equation}\label{eqn:spectral_clust_number}
	k = \underset{i}{\argmax} \big(|\varsigma_{i+1} - \varsigma_i|\big), \quad i=\Seta{|\Set B|},
\end{equation}
where $\varsigma_i$ is the $i$-th smallest eigenvalue of $\vect L$.

\subsection{Cluster-based coordination}

Clustering BSs serves two main purposes: {\it (i)} reducing intra-cluster interference and {\it (ii)} efficiently offload UEs from BSs which need to sleep to active BSs.
Interference reduction helps to increase the BS capacity in which BSs are capable to serve larger number of UEs. 
A BS can switch OFF if it does not serve UEs or it can offload the serving UEs to other BSs.
Therefore, BSs in a cluster have higher chance to switch OFF or to support the BSs who needs to be switched OFF.

Once the clusters are formed, the heavily loaded BS within the cluster is selected as the cluster head.
The function of a cluster head is to coordinate the transmissions among the cluster members by allocating orthogonal resource blocks between them in the time-domain.
Consequently, the entire load of the cluster is distributed between its members and orthogonal resource allocation helps to mitigate intra-cluster interference.
Due to fact that the BSs within a cluster have the ability to coordinate, the entire cluster can be seen as a single super BS which serves all the UEs within its vicinity as illustrated in Fig.~\ref{fig:coordination}.

Allowing UE offloading among BSs within the cluster helps to enable sleep mode for BSs while ensuring the UE satisfaction.
The offloading is carried out with the objective of minimizing the cluster load.
This reduces the number of UEs served with low rates.
Let $\Set{M}_{\Set{C}}$ be the set of UEs associated with the set of BSs in cluster $\Set{C}$.
Let $z_{bm}$ be the indicator which defines the connectivity between UE $m\in\Set{M}_{\Set{C}}$ and BS $b\in\Set{C}$, i.e. $z_{bm}=1$ if UE $m$ served by BS $b$ and, $z_{bm}=0$ otherwise. 
Thus, for a given set of BS transmission powers, the relaxed problem of \emph{mixed integer linear program} (MILP) for scheduling within the cluster $\Set{C}\in\overline{\Set{C}}$ is given by;
\begin{subequations}
\begin{eqnarray}\label{eqn:scheduling_relaxed}
	 \underset{\hat{\vect{z}}}{\text{minimize}} && \textstyle\sum_{\forall b\in\Set{C}} \rho_b \\
			\label{cns:unique_connection_new} \text{subject to} && \textstyle\sum_{\forall b\in\Set{C}_k} \hat{z}_{bm} = 1, \quad \forall m\in\Set{M}_{\Set{C}} \\
			 \label{cns:boolean_relaxed} && 0 \leq \hat{z}_{bm} \leq 1, \quad \forall m\in\Set{M}_{\Set{C}}, \forall b \in \Set{C}.
\end{eqnarray}
\end{subequations}
where $\hat{\vect{z}}$ is the relaxed optimization variable of the variables $z_{bm}$.
A suboptimal scheduling within the cluster is obtained by the optimal solution $(\hat{\vect{z}})^\star$ of (\ref{eqn:scheduling_relaxed}) as follows;
\begin{equation}
	(z_{bm})^\star =
	\begin{cases}
		1 & \mbox{if~} (\hat{z}_{bm})^\star = \argmax_{\forall b' \in \Set{C}}\Big( (z_{b'm})^\star \Big), \\
		0 & \mbox{otherwise}.
	\end{cases}
\end{equation}

\section{Self-Organizing Switching ON/OFF Mechanism}\label{sec:solution}

Given the formation of the clusters, our next goal is to propose a self-organizing solution for (\ref{eqn:optimization_energy_efficiency_network})-(\ref{cns:state}) in which each cluster of BSs individually adjusts its transmission parameters based on local information.
To do so, we use a regret-based learning approach~\cite{jnl:qian12}, in which the proposed solution consists of two interrelated parts: user association and cluster-wise BS transmission optimization.

\subsection{Load-Based User Association}\label{sec:ue_association_policy}

When a UE enters to the SCN or a UE does not satisfy regarding the service from currently serving BS, it seeks a new candidate BS who can ensure the quality of the service.
Therefore, a suitable UE association rule is required for UEs to make the above mention decision.
Classical UE association techniques based on received signal strength or signal to interference ratio  
are oblivious to the base stations' and the network's traffic load. 
This may lead to overloading BSs and lower spectral efficiencies.
Thus, a smarter mechanism in which the BSs advertise their load to all UEs within their coverage area is desirable.

At time instant $t$, each BS $b$ advertises its estimated load $\hat{\rho}_b(t)$ via a broadcast control message along with its transmission power $P_b(t)$. 
Considering both the received signal strength and load, at time $t$ the UE at location $\vectx$ connects to BS $b(\vectx,t)$, $\vectx\in\Set{L}_{b(\vectx, t)}$, according to the following UE association rule:
\begin{equation}\label{eqn:ue_association}
	b(\vectx, t) = \argmax_{b \in B} \Big\{ \Big(1-\hat{\rho}_b(t)\Big)^{\delta}P_b^{\texth{Rx}}(t) \Big\}.
\end{equation}
Here, $P_b^{\texth{Rx}}(t) =  P_b(t)I_b(t) h_{b}(\vectx,t)$ 
is the received signal power at the UE in location $\vectx$ from BS $b$ at time $t$.
The impact of the load is determined by the coefficient $\delta\geq 0$.
The classical reference signal strength indicator (RSSI)-based UE association is a special case of (11) when $\delta = 0$.
For $\delta>0$, as the BS load increases, the UEs are less likely to connect.

Due to fact that the BSs need to estimate their loads beforehand, the estimations must accurately reflect the actual load.
In order to obtain an accurate estimation for the load of the BS $b$, we compute the load estimation $\hat{\rho}_b(t)$ at time $t$ based on history as follows:
\begin{equation}\label{eqn:load_estimation}
	\hat{\rho}_b(t) = \hat{\rho}_b(t-1) + \nu(t) \Big( \rho_b(t-1)-\hat{\rho}_b(t-1) \Big),
\end{equation}
where $\nu(t)$ is the learning rate of the load estimation.
Leveraging different time-scales, we assume that $\nu(t)$ is selected such that the load estimation procedure~(\ref{eqn:load_estimation}) is much slower than the UE association process.

\subsection{Game Formulation}

\begin{table}[!t]
\caption{Simulation parameters.}
\vspace{-10pt}
\centering
\begin{tabular}{l c}
\hline
{\bf Parameter} & {\bf Value} \\
\hline \hline
Carrier frequency	& $2$ GHz \\
System bandwidth	& $10$ MHz \\
Thermal noise ($N_0$) &$-174$ dBm/Hz \\
Mean packet arrival rate $\big(\lambda(\vectx)/\mu(\vectx)\big)$ & 180 kbps\\
Maximum transmission powers: MBS, SBS & $46,~30$ dBm \\
\hline \multicolumn{2}{c}{\bf Minimum  distances} \\ \hline
MBS -- SBS, MBS -- UE & 75 m, 35m \\
SBS -- SBS, SBS -- UE & 40 m, 10 m \\
\hline \multicolumn{2}{c}{{\bf Path loss models} ($d$ in km)} \\ \hline
MBS - UE & $128.1 + 37.6\log_{10}(d)$\\
SBS - UE & $140.7 + 37.6\log_{10}(d)$\\
\hline \multicolumn{2}{c}{{\bf Clustering}} \\ \hline
Range of neighborhood ($\varepsilon_d$) & 250 m\\
Impact of neighborhood width ($\sigma_d, \sigma_l$) & 300, 1 \\
Tradeoff between similarities ($\theta$) & 0.5 \\
\hline \multicolumn{2}{c}{\bf Learning} \\ \hline
Boltzmann temperature ($\kappa$) & 10\\
Energy and load impacts on cost ($\alpha,~\beta$) & 0.5 W$^{-1}$,~0.5 \\
learning rate exponents for $\tau, \iota~\mbox{and}~\varepsilon$ & $0.6,~0.7,~0.8$ \\
\hline
\end{tabular}
\label{tab:sim_para}
\vspace{4pt}
\end{table}

In the proposed approach, with the knowledge of the scheduling within the cluster from (\ref{eqn:scheduling_relaxed}), the clusters need to autonomously select $\vect{v}$ in order to minimize their cost functions.
However, the cell coverages and the achievable throughputs of BSs depend not only on the action of their own cluster, but also affected by the choices of neighboring clusters due to the interference.
In this regard, we formulate a non-cooperative game $\Set{G} = \big( \overline{\Set{C}}, \{\Set{A}_\Set{C}\}_{\Set{C}\in\overline{\Set{C}}}, \{{u}_\Set{C}\}_{\Set{C}\in\overline{\Set{C}}} \big)$ in which the set of clusters ($\overline{\Set{C}}$) is the set of players.
Each player $\Set{C}\in\overline{\Set{C}}$ has a set $\Set{A}_{\Set{C}}=\big\{ a_{\Set{C},1},\ldots,a_{\Set{C},|\Set{A}_{\Set{C}}|} \big\}$ of actions where an action of cluster $\Set{C}$, $a_\Set{C}$, is composed of the configurations of all its cluster members, i.e. $a_\Set{C}=(v_{b_1},\ldots,v_{b_{|\Set{C}|}})$ with $\{b_1,\ldots,b_{|\Set{C}|}\}=\Set{C}$.
The load $\vect{\rho}$ of the system depends on the action $a_\Set{C}$ of cluster $\Set{C}$ and the actions of the other clusters $\vect{a}_{-\Set{C}}$.
$u_\Set{C}$ is the utility function of cluster $\Set{C}$ with $u_\Set{C}:\Set{A}_\Set{C}\rightarrow\mathbb{R}$ where $u_\Set{C}(a_\Set{C},\vect{a}_{-\Set{C}}) = -\Gamma_\Set{C}(\vect{\rho})$.
Let $\vect{\pi}_\Set{C}(t) = \big[ \pi_{\Set{C},1}(t),\ldots,\pi_{\Set{C},|\Set{A}_\Set{C}|}(t) \big]$ be a probability distribution using which cluster $\Set{C}$ selects a given action from $\Set{A}_\Set{C}$ at time instant $t$.
Here, $\pi_{\Set{C},i}(t)=\mbox{Pr}\big(a_\Set{C}(t)=a_{\Set{C},i}\big)$ is cluster $\Set{C}$'s \emph{mixed strategy} where $a_\Set{C}(t)$ is the action of player $\Set{C}$ at time $t$.
While cluster $\Set{C}$ plays action $a_\Set{C}(t)$, it may regret or be satisfied about the action it played based on the observed utility feedback ${u}_\Set{C}(t)$. 
Therefore, player $\Set{C}$ estimates its utility ${\vect{\hat{u}}}_\Set{C}(t)=\big[ \hat{u}_{\Set{C},1}(t),\ldots,\hat{u}_{\Set{C},|\Set{A}_\Set{C}|}(t) \big]$ and regret ${\vect{\hat{r}}}_\Set{C}(t)=\big[ \hat{r}_{\Set{C},1}(t),\ldots,\hat{r}_{\Set{C},|\Set{A}_b|}(t) \big]$ for each action assuming it has played the same action during all previous time slots $\{\Seta{t-1}\}$.
At each time $t$, player $\Set{C}$ updates its mixed strategy probability distribution $\vect{\pi}_\Set{C}$ in which the actions with higher regrets are exploited while exploring the actions with low regrets. 
Such behavior can be captured by the Boltzmann-Gibbs (BG) distribution $\vect{G}_\Set{C}=(G_{\Set{C},1},\ldots,G_{\Set{C},|\Set{A}_\Set{C}|})$ which is calculated as follows:
\begin{equation}
	G_{\Set{C},i}\big(\vect{\hat{r}}_b(t)\big) = \frac{\exp\big(\kappa_\Set{C} \hat{r}_{\Set{C},i}^+(t)\big)} {\sum_{\forall i'\in\Set{A}_\Set{C}} \exp\big(\kappa_\Set{C} \hat{r}_{\Set{C},i'}^+(t)\big) }, \: i\in\Set{A}_\Set{C},
\end{equation}
where $\kappa_\Set{C}>0$ is a temperature parameter which balances between exploration and exploitation.
For each time $t$, all the estimations for any player $\Set{C}\in\overline{\Set{C}}$, $\vect{\hat{u}}_\Set{C}(t),~\vect{\hat{r}}_\Set{C}(t)$ and $\vect{\pi}_\Set{C}(t)$, are updated as follows;
\begin{equation}\label{eqn:algoUpdates}
\begin{cases}
	{\hat{u}}_{\Set{C},i}(t) &= {\hat{u}}_{\Set{C},i}(t-1)  \\
	&\hfill + \tau_\Set{C}(t)\mathds{1}_{\{a_{\Set{C},i}=v_\Set{C}(t-1)\}} \Bigl({u}_\Set{C}(t)-{\hat{u}}_{\Set{C},i}(t-1)\Bigr),\\
	{\hat{r}}_{\Set{C},i}(t) &= {\hat{r}}_{\Set{C},i}(t-1)  \\
	&\hfill +\iota_\Set{C}(t) \Bigl({\hat{u}}_{\Set{C},i}(t-1)-{u}_\Set{C}(t-1)-{\hat{r}}_{\Set{C},i}(t-1)\Bigr),\\
	\pi_{\Set{C},i}(t) &= \pi_{\Set{C},i}(t-1) \\
	&\hfill + \varepsilon_\Set{C}(t) \Bigl(G_{\Set{C},i}\big({\vect{\hat{r}}}_\Set{C}(t-1)\big)-\pi_{\Set{C},i}(t-1)\Bigr).
\end{cases}
\end{equation}
\normalsize
with the learning rates satisfying $\lim_{t\rightarrow\infty}\sum_{n=1}^t\xi(n) = \infty$ and $\lim_{t\rightarrow\infty}\sum_{n=1}^t\xi^2(n) < \infty$ for all $\xi=\{\tau,\iota,\varepsilon\}$.
This process guarantees the convergence of the algorithm to an $\epsilon$-coarse correlated equilibrium~\cite{jnl:qian12}.
Our choice of the learning rates follows the format of ${1}/{t^\phi}$ with exponent $\phi\in(0.5,1)$.

At the beginning of each time instant $t$, all BSs advertise their loads $(\hat{\rho}_b)$ and the clusters select their actions $(a_\Set{C})$ based on their mixed-strategy probabilities $(\pi_\Set{C})$.
Based on the estimated load, UEs associate as per $a_\Set{C}(t)$.
All the BSs carry out the transmission based on the actions $\big(a_1(t),\ldots,a_{|\overline{\Set{C}}|}(t)\big)$ and calculate the utilities $u_\Set{C}(t)$.
Each cluster individually updates its utility and regret estimations $\big(\vect{\hat{u}}_\Set{C}(t),\vect{\hat{r}}_\Set{C}(t)\big)$ along with the mixed strategy probabilities $\big(\pi_\Set{C}(t)\big)$.

\section{Simulation Results}\label{sec:results}

For simulations, we consider a single cell covered by a macro BS with an arbitrary number of SBSs and UEs uniformly distributed over the area.
All the BSs share the entire spectrum and thus, suffer co-channel interference.
We conduct multiple simulations for various practical configurations and the presented results are averaged over a large number of runs.
The parameters used for the simulations are summarized in Table~\ref{tab:sim_para}.
The proposed cluster-based coordination and learning based sleep-wake mechanisms is compared with the conventional network operation referred to hereinafter as ``classical approach'' in which BSs do not have the capability to switch between sleep-wake states.
For further comparisons, we consider a system of BSs which uses the learning based sleep-wake mechanism for implicit coordination without forming clusters.
It is referred to as ``learning approach without clusters" throughout the rest of the discussion.

\begin{figure}[!t]
\centering
\includegraphics[trim = 0mm 0mm 0mm 5mm, clip,keepaspectratio,width=\myfigfactor\textwidth]{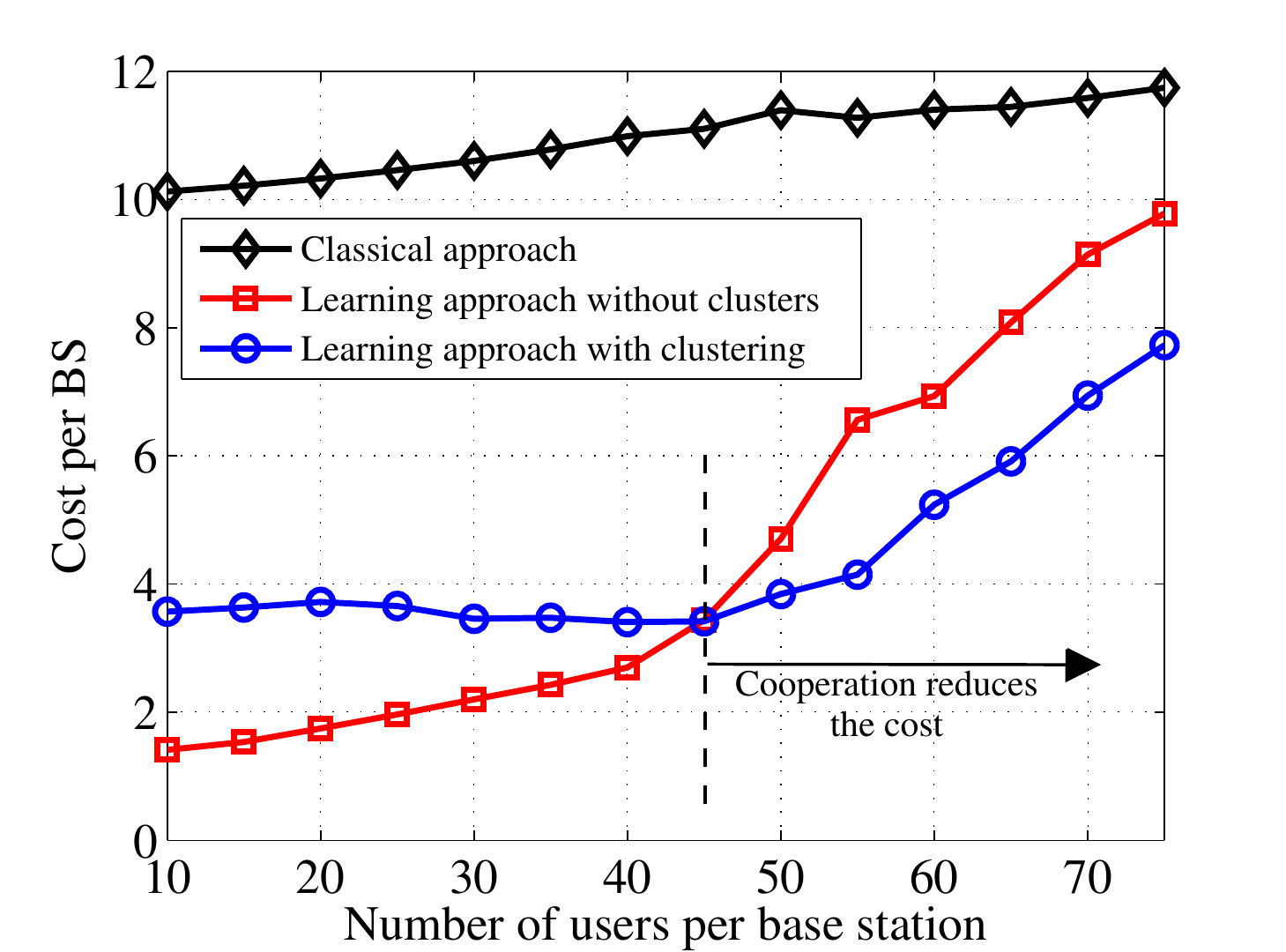}
\vspace{-8pt}
\caption{Variation of the cost per BS as a function of the number of UEs with 10 SBSs. The average number of clusters in the system is $5$ and the joint similarity is used for the clustering.}
\label{fig:cost_vs_ue}
\vspace{-5pt}
\end{figure}

Fig. \ref{fig:cost_vs_ue} shows the changes in the average cost per BS as the number of UEs varies.
As the number of UEs increases, the load in the system as well as the BS energy consumption increases and thus, the average cost increases.
However, Fig~\ref{fig:cost_vs_ue} shows that the proposed learning approaches can reduce the cost by balancing the energy consumption and the handled load.
In Fig.~\ref{fig:cost_vs_ue}, we can see that for small network sizes, the cluster-based approach consumes more energy than the learning approach without clustering. 
This is due to the fact that, for such networks, only a small number of BSs needs to be active and, in such a case, the cluster-based approach would require extra energy for coordination.
In contrast, for highly-loaded networks, as seen in Fig.~\ref{fig:cost_vs_ue}, clustering allows to better offload traffic and, subsequently, improve the overall energy efficiency of the system. 
In this respect, Fig.~\ref{fig:cost_vs_ue} shows that clustering yields up to $47 \%$ and $25 \%$ of cost reduction, relative to the classical approach and to learning without clusters, for a network with 65 UEs.

\begin{figure}[!t]
\centering
\includegraphics[trim = 0mm 0mm 0mm 5mm, clip,keepaspectratio,width=\myfigfactor\textwidth]{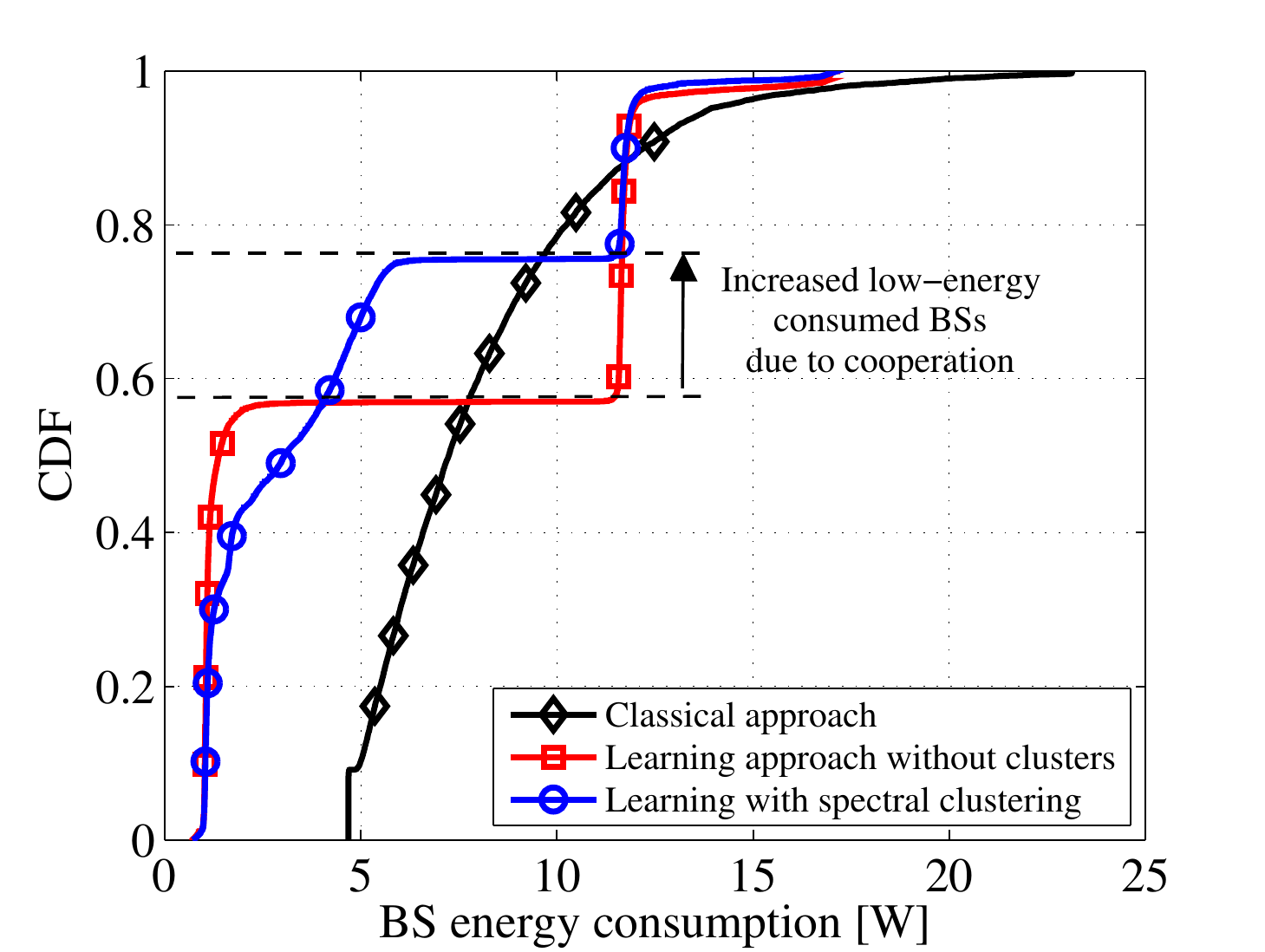}
\vspace{-10pt}
\caption{Cumulative density function of BS energy consumption for 10 SBSs and 10 -- 75 UEs. The average number of the cluster in the system is $5$ and the joint similarity is used for the clustering.}
\label{fig:cdf_energy}
\end{figure}

In Fig. \ref{fig:cdf_energy}, we show the CDF of the BSs' energy consumption.
First, we can see that for the classical approach, the frequency of BSs having a high energy consumption is much higher than in the cases with learning.
Indeed, the proposed learning method allows lightly-loaded BSs to offload their traffic and switch to sleep mode, thus yielding significant energy reductions.
Coordination between clusters allows more BSs to switch to sleep mode and, thus, as shown in Fig. 3,  the proposed learning approach with clustering can yield a larger number of BSs who consume less energy.
The average energy consumption reductions with the cluster-based approach are 40\% and 17\% compared to the classical approach and the learning approach without clusters, respectively.

\begin{figure}[t]
\centering
\includegraphics[trim = 0mm 0mm 0mm 5mm, clip,keepaspectratio,width=\myfigfactor\textwidth]{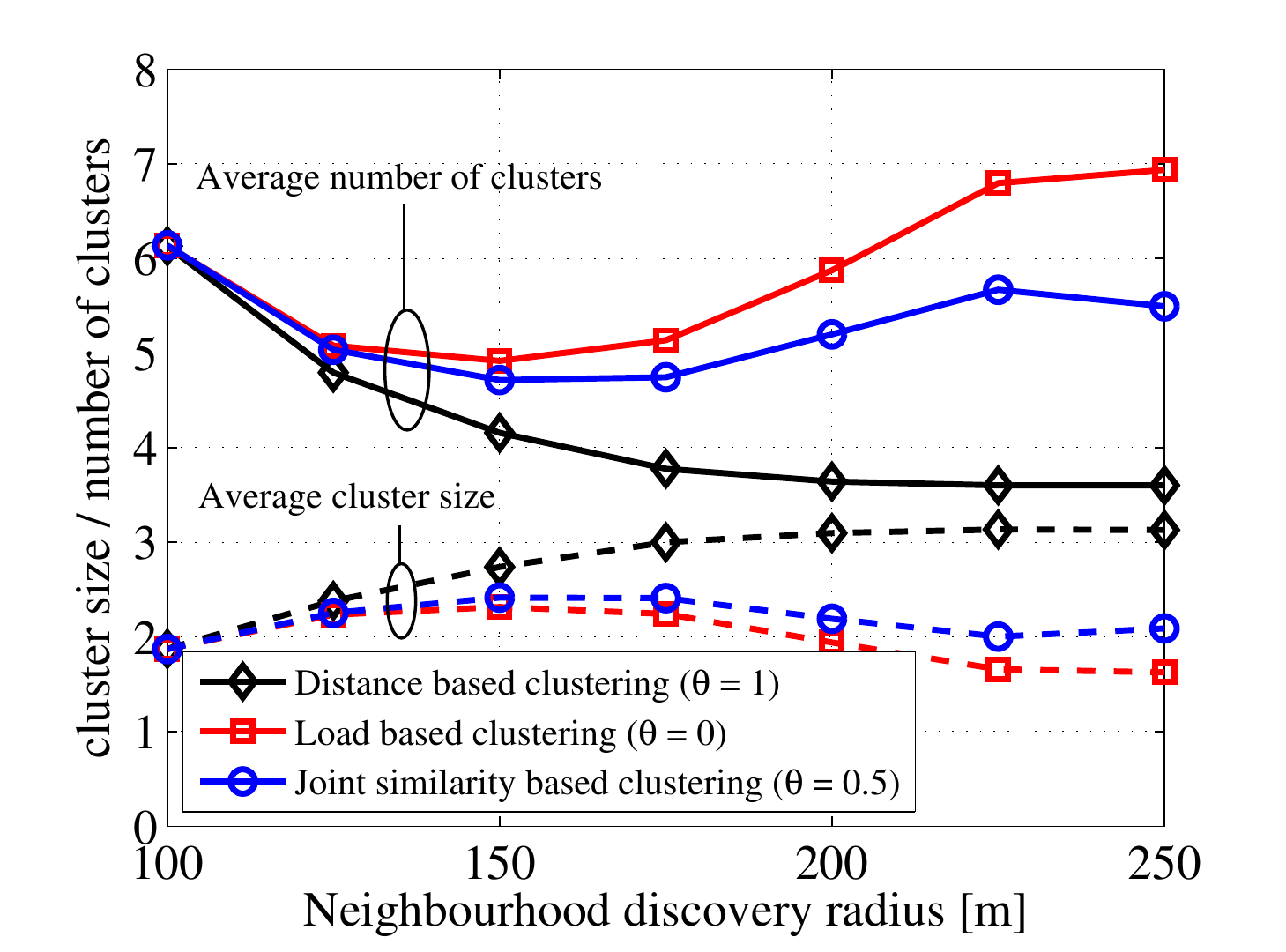}
\vspace{-10pt}
\caption{Comparison of the average number of clusters and the average cluster size with the similarities calculated for $\theta = \{0, 0.5, 1\}$. The setup is for 10 SBSs and 50 UEs.}
\label{fig:clusters_comparison}
\end{figure}

Fig.~\ref{fig:clusters_comparison} shows the average number of clusters and the average cluster sizes for distance-based, load-based and joint similarities.
As the neighborhood discovery radius ($\varepsilon_d$) increases, the non-zero edges in $G=(\Set B, \Set E)$ increases.
Therefore, clustering based on distance similarity allows to group more BSs together thus yielding a smaller number of clusters with large cluster size.
As the cluster size increases, the cluster load increases which directly impacts the network cost.
As shown in Fig.~\ref{fig:clusters_comparison}, for $\theta<1$, up to $\varepsilon_d<175$, the network cost is reduced by forming large clusters.
However, for $\varepsilon_d>175$, increasing the load affects the similarity and thus, forming large clusters is avoided.

\section{Conclusions}\label{sec:conclusions}

In this paper, we have proposed a cluster-based opportunistic on/off mechanism for small cell base stations.
We have formulated the problem as a non-cooperative game in which the goal of each cluster of base stations is to minimize the cluster cost which captures the energy and load expenditures.
To solve the game, we have proposed a distributed algorithm and an intra-cluster coordination method using which the base stations choose their transmission modes with little additional overhead. 
Our proposed clustering method uses the information on both BSs locations and their capability of handling loads in order to improve the overall performance.
Simulation results have shown significant gains in energy expenditure and load reduction compared to the conventional transmission techniques.
Moreover, the results provide an insight of when and how to reap the benefits from the cluster-based coordination in small cell networks.

\bibliographystyle{IEEEtran}
\bibliography{IEEEabrv}

\begin{thebibliography}{10}
\providecommand{\url}[1]{#1}
\csname url@samestyle\endcsname
\providecommand{\newblock}{\relax}
\providecommand{\bibinfo}[2]{#2}
\providecommand{\BIBentrySTDinterwordspacing}{\spaceskip=0pt\relax}
\providecommand{\BIBentryALTinterwordstretchfactor}{4}
\providecommand{\BIBentryALTinterwordspacing}{\spaceskip=\fontdimen2\font plus
\BIBentryALTinterwordstretchfactor\fontdimen3\font minus
  \fontdimen4\font\relax}
\providecommand{\BIBforeignlanguage}[2]{{%
\expandafter\ifx\csname l@#1\endcsname\relax
\typeout{** WARNING: IEEEtran.bst: No hyphenation pattern has been}%
\typeout{** loaded for the language `#1'. Using the pattern for}%
\typeout{** the default language instead.}%
\else
\language=\csname l@#1\endcsname
\fi
#2}}
\providecommand{\BIBdecl}{\relax}
\BIBdecl

\bibitem{pap:brevis11}
P.~Gonzalez-Brevis, J.~Gondzio, Y.~Fan, H.~V. Poor, J.~Thompson, I.~Krikidis,
  and P.-J. Chung, ``Base station location optimization for minimal energy
  consumption in wireless networks,'' in \emph{Proc. IEEE VTC}, May 2011, pp.
  1--5.

\bibitem{jnl:amr14}
A.~Abdelnasser, E.~Hossain, and D.~I. Kim, ``Clustering and resource allocation
  for dense femtocells in a two-tier cellular ofdma network,'' \emph{{IEEE}
  Trans. Wireless Commun.}, vol.~13, no.~3, pp. 1628--1641, Mar. 2014.

\bibitem{pap:bhuaumik10}
\BIBentryALTinterwordspacing
S.~Bhaumik, G.~J. Narlikar, S.~Chattopadhyay, and S.~Kanugovi, ``Breathe to
  stay cool: adjusting cell sizes to reduce energy consumption,'' in
  \emph{Proc. ACM SIGCOMM on Green Netw.}, ser. Green Networking '10, 2010, pp.
  41 -- 46. [Online]. Available:
  \url{http://doi.acm.org/10.1145/1851290.1851300}
\BIBentrySTDinterwordspacing

\bibitem{pap:soh13}
Y.~S. Soh, T.~Q. Quek, and M.~Kountouris, ``Dynamic sleep mode strategies in
  energy efficient cellular networks,'' in \emph{Proc. IEEE ICC}, Jun. 2013,
  pp. 3131--3136.

\bibitem{pap:hoisseini12}
K.~Hosseini, H.~Dahrouj, and R.~Adev, ``Distributed clustering and interference
  management in two-tier networks,'' in \emph{Proc. IEEE GLOBECOM}, Dec. 2012,
  pp. 4267--4272.

\bibitem{pap:lee13}
G.~Lee, H.~Kim, Y.-T. Kim, and B.-H. Kim, ``Delaunay triangulation based green
  base station operation for self organizing network,'' in \emph{Proc. IEEE
  GreenCom-iThings-CPSCom}, Aug. 2013, pp. 1--6.

\bibitem{jnl:luxburg07}
\BIBentryALTinterwordspacing
U.~von Luxburg, ``A tutorial on spectral clustering,'' \emph{CoRR}, vol.
  abs/0711.0189, 2007. [Online]. Available:
  \url{http://arxiv.org/abs/0711.0189}
\BIBentrySTDinterwordspacing

\bibitem{jnl:qian12}
L.~P. Qian, Z.~Y.J.A., and M.~Chiang, ``Distributed nonconvex power control
  using {G}ibbs sampling,'' \emph{{IEEE} Trans. Commun.}, vol.~60, no.~12, pp.
  3886--3898, 2012.

\bibitem{jnl:kyuho11}
K.~Son, H.~Kim, Y.~Yi, and B.~Krishnamachari, ``Base station operation and user
  association mechanisms for energy-delay tradeoffs in green cellular
  networks,'' \emph{{IEEE} J. Sel. Areas Commun.}, vol.~29, no.~8, pp.
  1525--1536, Sep. 2011.

\bibitem{pap:frenger11}
P.~Frenger, P.~Moberg, J.~Malmodin, Y.~Jading, and I.~Godor, ``Reducing energy
  consumption in lte with cell dtx,'' in \emph{Proc. IEEE VTC}, May 2011, pp.
  1--5.

\bibitem{jnl:hongseok12}
H.~Kim, G.~de~Veciana, X.~Yang, and M.~Venkatachalam, ``Distributed
  $\alpha$-optimal user association and cell load balancing in wireless
  networks,'' \emph{{IEEE/ACM} Trans. Netw.}, vol.~20, no.~1, pp. 177--190,
  Feb. 2012.

\end{thebibliography}

\end{document}